\documentclass[12pt]{article}
\usepackage{times}
\usepackage{geometry}
\usepackage{amssymb}   
\usepackage{hyperref}  
\usepackage{ragged2e}
\usepackage[figure,figure*]{hypcap}
\usepackage{ulem} 
\usepackage[T1]{fontenc}
\usepackage{graphicx}
\usepackage{caption}
\usepackage{subcaption}
\usepackage[table]{xcolor}

\usepackage{color}
\usepackage{lineno}
\usepackage{wrapfig}
\usepackage{subcaption}
\usepackage{float}

\usepackage[utf8]{inputenc}
\usepackage{enumitem,amssymb}
\usepackage{ragged2e}

\newlist{thematic}{itemize}{8}
\setlist[thematic]{label=$\square$}
\usepackage{pifont}

\begin{document}
\pagestyle{empty}
\raggedright
\huge
Astro2020 APC White Paper \linebreak

Another Servicing Mission to Extend Hubble Space Telescope's Science past the Next Decade
\linebreak
\normalsize

\noindent \textbf{Thematic Areas:} \hspace*{20pt} $\boxtimes$ Activity \hspace*{20pt} $\square$ Project \hspace*{20pt} $\square$ State of the Profession

\textbf{Principal Author:}

Name: Mercedes L\'opez-Morales	
 \linebreak						
Institution: Center for Astrophysics | Harvard \& Smithsonian 
 \linebreak
Email: mlopez-morales@cfa.harvard.edu
 \linebreak
Phone: +1.617.496.7818
 \linebreak
 
\textbf{Co-authors:} 
Kevin France (University of Colorado), Francesco R. Ferraro (University of Bologna), Rupali Chandar (University of Toledo), Steven Finkelstein (University of Texas at Austin), Stephane Charlot (Institut d'Astrophysique de Paris - CNRS/Sorbonne Université), Gilda Ballester (University of Arizona), Melina. C. Bersten (Instituto de Astrof\'isica de La Plata CONICET-UNLP; University of Tokyo), Jose M. Diego (Instituto de Fisica de Cantabria), Gast\'on Folatelli (Instituto de Astrof\'isica de La Plata CONICET-UNLP; University of Tokyo), Domingo Garc\'ia-Senz (Universitat Polit\'ecnica de Catalunya), Mauro Giavalisco (University of Massachusetts), Rolf A. Jansen (Arizona State University),  Patrick L. Kelly (University of Minnesota), Thomas Maccarone (Texas Tech University), Seth Redfield (Wesleyan University),  Pilar Ruiz-Lapuente (University of Barcelona), Steve Shore (University of Pisa), Nitya Kallivayalil (University of Virginia) 
\\
\vspace{0.4cm}

\textbf{Co-signers:} 
Munazza K. Alam (Center for Astrophysics | Harvard \& Smithsonian), Juan Manuel Alcal\'a (INAF – AO of Naples), Jay Anderson (Space Telescope Space Institute), Daniel Angerhausen (CSH, Bern University), Francesca Annibali (INAF – OAS of Bologna), D\'aniel Apai (University of Arizona),  David Ardila (Jet Propulsion Laboratory), Santiago Arribas (Centro de Astrobiolog{\'\i}a CSIC-INTA), Hakim Atek (Institut d'astrophysique de Paris - CNRS/Sorbonne Université), Thomas R. Ayres, (University of Colorado), Francesca Bacciotti (INAF – OA of  Florence), Beatriz Barbuy (Universidade de Sao Paulo), Joanna K. Barstow (University College London), Nate Bastian (Liverpool John Moores University), Natasha E. Batalha (University of California Santa Cruz), Matthew Bayliss (Kavli Institute for Astrophysics and Space Research | MIT), Jacob L.\ Bean (University of Chicago), Giacomo Beccari (European Southern Observatory), Tracy M. Becker (Southwest Research Institute), Peter Behroozi (University of Arizona), Michele Bellazzini (INAF –OAS Bologna), Andrea Bellini (Space Telescope Science Institute), Björn Benneke (University of Montreal), Danielle A. Berg (The Ohio State University), Marco Berton (University of Turku), Sergi Blanco-Cuaresma (Center for Astrophysics | Harvard \& Smithsonian), Bertrand Bonfond (Université de Liege), Stefano Borgani (University of Trieste), Vincent Bourrier (Universit\'e de Gen\`eve), Angela Bragaglia (INAF – OAS Bologna), Jonathan Brande (NASA GSFC, University of Maryland),Giovanni Bruno (INAF - OA Catania), Andy Bunker (University of Oxford), Joseph N. Burchett (UC - Santa Cruz), Ivan Cabrera-Ziri (Center for Astrophysics | Harvard \& Smithsonian), Mario Cadelano (University of Bologna), Nelson Caldwell (Center for Astrophysics | Harvard \& Smithsonian), Aarynn Carter (University of Exeter), Salvatore Capozziello (University of Naples), Roberto Capuzzodolcetta (University of La Sapienza), Ludmila Carone (Max-Planck-Institut f\"ur Astronomie), Santi Cassisi (INAF- OA of Abruzzo), Marco Castellano (INAF- OA of Rome), Jessie L. Christiansen (Caltech/IPAC-NExScI), Andrea Cimatti (University of Bologna), Paula Coelho (Universidade de Sao Paulo), Karen Collins (Center for Astrophysics | Harvard \& Smithsonian) Knicole Colon (NASA Goddard), Thierry Contini (IRAP, Toulouse), Enrico Corongiu (Las Campanas Observatory), Enrico Maria Corsini (University of  Padua), Nicolas B. Cowan (McGill University), Denija Crnojevi\'c (University of Tampa), H{\aa}kon Dahle (University of Oslo), Emanuele Dalessandro (INAF-OAS Bologna), Elena Dalla Bont\`a (University of Padua), Daniele Dallacasa (University of Bologna), Filippo Dammando (INAF-IRA Bologna), Stefano Oliveira de Souza (Universidade de Sao Paulo), Scilla Degl’innocenti (University of Pisa), Domitilla De Martino (INAF – OA Naples), Andr\'es del Pino (Space Telescope Science Institute), Antonaldo Diaferio (University of Turin), Tiziana Di Salvo (University of Palermo), Chuanfei Dong (Princeton University), Leonardo A. dos Santos (Universit\'e de Gen\`eve), Andrew Dolphin (Raytheon \& University of Arizona), Aaron Dotter (Center for Astrophysics | Harvard \& Smithsonian), William Dunn (UCL-MSSL), Andrea Dupree (Center for Astrophysics | Harvard \& Smithsonian), David Ehrenreich (Universit\'e de Gen\`eve), David Elbaz (CEA Saclay), Sylvia Ekstr\"om (Universit\'e de Gen\`eve), Catherine Espaillat (Boston University), N\'estor Espinoza (Max-Planck-Institut f\"ur Astronomie), Alexander Fritz (INAF-IASF Milano), Adriano Fontana (INAF – OA Rome), Luigi Foschini (INAF- OA of Brera), Luca Fossati (Austrian Academy of Sciences), Thierry Fouchet (Observatoire de Paris), Marijn Franx (University of Leiden), Cynthia S.\ Froning (University of Texas at Austin), Flavio Fusi Pecci  (INAF –OAS Bologna), Carmen Gallart (Istituto de Astrofisica de Canarias), Giuseppe Galletta (University of Padua), Davide Gandolfi (University of Turin), Peter Gao (University of California Berkeley), Miriam Garcia (Centro de Astrobiolog\'ia, CSIC-INTA), Antonio Garc\'ia Mu\~noz (Technische Universit\"at Berlin), John E. Gizis (University of Delaware), Jayesh M. Goyal (University of Exeter), Or Graur (Center for Astrophysics | Harvard \& Smithsonian), Maximilian N. G{\"u}nther (Massachusetts Institute of Technology), Nimish Hathi (Space Telescope Science Institute), Craig Heinke (University of Alberta), David Hendel (University of Toronto), Gregory W. Henry (Tennessee State University), Gregory Herczeg (Kavli Institute for Astronomy and Astrophysics, Peking University), Ricardo Hueso (University of the Basque Country, UPV/EHU), Sean D. Johnson (Princeton University \& Carnegie Observatories),  
Lisa Kaltenegger (Cornell University), Margarita Karovska (Center for Astrophysics | Harvard \& Smithsonian), Jeyhan Kartaltepe (Rochester Institute of Technology), Eliza Kempton (University of Maryland), John Kielkopf (University of Louisville), James Kirk (Center for Astrophysics | Harvard \& Smithsonian), Edwin S. Kite (University of Chicago), Heather A. Knutson (California Institute of Technology), Laura Kreidberg (Center for Astrophysics | Harvard \& Smithsonian), Fabio La Franca (University of Roma Tre), Alvaro Labiano (Centro de Astrobiolog\'ia CSIC-INTA), Laurent Lamy (LESIA, Observatoire de Paris, PSL, CNRS), Lauranne Lanz (Dartmouth College), Barbara Lanzoni (University of Bologna), Soeren Larsen (Radboud University Nijmegen), Baptiste Lavie (Universit\'e de Gen\`eve), Charles J. Law (Center for Astrophysics | Harvard \& Smithsonian), Vianney Lebouteiller (AIM, CEA Saclay, France), Alain Lecavelier des Etangs (Institut d'astrophysique de Paris-CNRS), J\'er\'emy Leconte (Laboratoire d'Astrophysique de Bordeaux | CNRS), Olivier Le~F\`evre (Laboratoire d’Astrophysique de Marseille - CNRS/Aix Marseille Université), Nicolas Lehner (University of Notre Dame), Matthew D. Lehnert  (Institut d'Astrophysique de Paris - CNRS/Sorbonne Universit\'e), Monika Lendl (Universit\'e de Gen\`eve), Nikole Lewis (Cornell University), Mattia Libralato (Space Telescope Science Institute), Jorge Lillo-Box (Center for Astrobiology, CAB) Michael R. Line (Arizona State University), Eric Lopez (NASA Goddard), Joshua Lothringer (University of Arizona), R. O. Parke Loyd (Arizona State University), Meredith A. MacGregor (Carnegie DTM), Jes\'us Ma{\'\i}z Apell\'aniz (Centro de Astrobiolog{\'\i}a CSIC-INTA), Carlo F. Manara (European Southern Observatory), Luigi Mancini (University of Rome Tor Vergata), Filippo Mannucci (INAF-AO of Arcetri), Megan Mansfield (University of Chicago), Massimo Marengo (Iowa State University), Paola Marigo (University of Padua), 
Francesca Matteucci (University of Trieste), Elena Mason (INAF – OA of Trieste), Davide Massari (University of Bologna), Pasquale Mazzotta (University of Rome Tor Vergata), Edward McClain (Center for Astrophysics | Harvard \& Smithsonian), Chima McGruder (Center for Astrophysics | Harvard \& Smithsonian), Kristen B.W. McQuinn (Rutgers University), Marco Merafina (University of Rome “La Sapienza”), Roberto Mignani (INAF-IASF Milan), Thomas Mikal-Evans (Kavli Institute for Astrophysics and Space Research | MIT), Antonino P. Milone (University of Padova),  Karan Molaverdikhani (Max Planck Institute for Astronomy), Alessio Mucciarelli (University of  Bologna), Rohan P. Naidu (Center for Astrophysics | Harvard \& Smithsonian), Giampiero Naletto (University of Padua), Domenico Nardiello (University of Padua), David Nataf (Johns Hopkins University), Roberto Nesci (INAF/IASPS Rome), Nikolay Nikolov (Johns Hopkins University), Mario Nonino (INAF – OA Trieste), Achille Nucita (University of Salento), Livia Origlia (INAF-OAS Bologna), Sergio Ortolani (University of Padua), Luisa Ostorero (University of Turin), Cristina Pallanca (University of Bologna), Vivien Parmentier (University of Oxford), Jay Pasachoff (Williams College—Hopkins Observatory), George G. Pavlov (Pennsylvania State University), Sibilla Perina (INAF-AO of Turin), Marshall D. Perrin (Space Telescope Science Institute), Giampaolo Piotto (University of Padua), Valerio Pirronello (University of Catania), Lucia Pozzetti (INAF – OAS Bologna), Benjamin V. Rackham (University of Arizona), Marc Rafelski (Space Telescope Science Institute), Emily Rauscher (University of Michigan),
Joseph Ribaudo (Providence College), R. Michael Rich (University of California Los Angeles), Johan Richard (Centre de Recherche Astrophysique de Lyon - CNRS/UCBL/ENS Lyon), Harvey Richer (University of British Columbia), Reed Riddle (CalTech), Liliana Rivera Sandoval (Texas Tech University), Giulia Rodighiero (University of Padua), Joseph E. Rodriguez (Center for Astrophysics | Harvard \& Smithsonian), Giuseppe Romeo (INAF- OA Catania), Michael J. Rutkowski (Minnesota State University), Maurizio Salaris (Liverpool John Moores University), Agust\'in S\'anchez-Lavega (University of the Basque Country, UPV/EHU), Nicoletta Sanna (INAF – OA of Florence), Ata Sarajedini (Florida Atlantic University), Riccardo Schiavon (Liverpool John Moores University), Everett Schlawin (University of Arizona), Daniel Schaerer (University of Geneva \& CNRS), Eric M. Schlegel (University of Texas San Antonio), Giancarlo Setti (Accademia dei Lincei), Keren Sharon (University of Michigan), Evgenya Shkolnik (Arizona State University), Joshua D. Simon (Carnegie Observatories), David K. Sing (Johns Hopkins University), John Southworth (Keele University), Alessandro Spagna (INAF – AO of Turin), Ian W. Stephens (Center for Astrophysics | Harvard \& Smithsonian), Kevin Stevenson (Space Telescope Science Institute), Andrew Szentgyorgyi (Center for Astrophysics | Harvard \& Smithsonian), Sandro Tacchella (Center for Astrophysics | Harvard \& Smithsonian), Jake Taylor (University of Oxford), Johanna Teske (Carnegie Observatories), Kamen O. Todorov (University of Amsterdam), Monica Tosi (INAF – OAS Bologna), Tommaso Treu (University of California Los Angeles), Todd Tripp (University of Massachusetts-Amherst), Elena Valenti (European Southern Observatory), Olivia Venot (Laboratoire Interuniversitaire des Syst\`emes Atmosph\'eriques | CNRS), Anne Verbiscer (University of Virginia), Enrico Vesperini (Indiana University), Alfred Vidal-Madjar (Institut d'astrophysique de Paris-CNRS), Angeles Perez Villegas (Universidade de Sao Paulo), Saeqa D. Vrtilek (Center for Astrophysics | Harvard \& Smithsonian), Hannah R. Wakeford (Space Telescope Science Institute), Q. Daniel Wang (University of Massachusetts), Ian C. Weaver (Center for Astrophysics | Harvard \& Smithsonian), Andrew Wetzel (University of California, Davis), Benjamin F. Williams (University of Washington), Jennifer G. Winters (Center for Astrophysics | Harvard \& Smithsonian), Julien de Wit (MIT), Aida Wofford (Universidad Nacional Autonoma de Mexico), Allison Youngblood (NASA Goddard), Manuela Zoccali (Pontificia Universitad Catolica de Chile), David R. Zurek (American Museum of Natural History), Robert T. Zellem (Jet Propulsion Laboratory - California Institute of Technology)
\justify

\vspace{0.5cm}
\textbf{Abstract}
 The Hubble Space Telescope has produced astonishing science over the past thirty years. Hubble's productivity can continue to soar for years to come provided some worn out components get upgraded. While powerful new ground-based and space telescopes are expected to come online over the next decade, none of them will have the UV capabilities that make Hubble a unique observatory. Without Hubble, progress in UV and blue optical astrophysics will be halted. Observations at these wavelengths are key for a range of unresolved astrophysics questions, ranging from the characterization of solar system planets to understanding interaction of galaxies with the intergalactic medium and the formation history of the universe. Hubble will remain our only source of high-angular resolution UV imaging and high-sensitivity UV spectroscopy for the next two decades, offering the ability for continued unique science and maximizing the science return from complementary observatories. Therefore, we recommend that NASA, ESA, and the private sector study the scientific merit, technical feasibility, and risk of a new servicing mission to Hubble to boost its orbit, fix aging components and expand its instrumentation. Doing so would: 1) keep Hubble on its path to reach its 
 full potential, 2) extend the mission's lifetime past the next decade, which will maximize the synergy of Hubble with other upcoming facilities, and 3) enable and enhance the continuation of scientific discoveries in UV and optical astrophysics.
 


\pagebreak
\pagestyle{plain}
\pagenumbering{arabic} 

\noindent {\bf Background}
\vspace{1mm}

In its almost 30 years of operations the Hubble Space Telescope (Hubble) has arguably proven to be the most important single scientific instrument ever developed. Some of Hubble's remarkable science results include mapping the expansion of the Universe, studying 
some of the earliest massive galaxies to form in the Universe,  discovering that all major galaxies harbor a super-massive black hole, unveiling how stars are born, and most recently probing the atmospheres of planets in our Solar System and orbiting other stars, and detecting light from a gravitational waves source.

Part of Hubble's longtime success comes from the series of servicing missions to repair parts degraded over time by the harsh space environment, and to install new enhanced instruments. As described in Table~1, Hubble has had five servicing missions between 1993 and 2009\footnote{https://www.nasa.gov/mission$\_$pages/hubble/servicing/index.html}. During those servicing missions astronauts compensated for Hubble's primary mirror aberration, refurbished and fixed existing instruments,  installed new ones, and replaced/installed other telescope components, such as  solar arrays, batteries, and gyroscopes. Owing to the retirement of the shuttle fleet, the fifth servicing mission in 2009 (SM4) was considered Hubble's last. One of the tasks performed during SM4 was to attach a device to the base of the telescope to facilitate de-orbiting when Hubble is eventually decommissioned. 

Over Hubble's lifetime so far, more than 20,000 individual astronomers worldwide have used Hubble data in their publications, with the telescope now producing almost 1000 refereed publications per year (see Fig.~1). By 2018, Hubble had produced over 16,000 refereed publications, which accumulated more than 800,000 citations\footnote{Data and figure obtained from http://archive.stsci.edu/Hubble/bibliography/pubstat.html}. Hubble data have also supported more than 600 Ph.D. theses. Telescope over-subscription rate (i.e. the fraction of proposals requesting telescope time that gets awarded) has been about 1:6 over the last five proposal cycles (Cycles 21-26).  In the most recent Cycle 27 over subscription rate jumped to 1:9, which shows the ongoing increasing demand for Hubble time. The demand for Hubble observing time and its productivity is at an all-time high, and demand for Hubble observations is expected to increase even more after the launch of the James Webb Space Telescope (JWST), for which Hubble can provide complementary key observations at UV and blue optical wavelengths that JWST cannot reach.

With its current suite of instruments, Hubble can observe at wavelengths from the FUV to the near-IR (0.1$\mu$m - 1.7$\mu$m). Having observations of Hubble overlap with other facilities that include coverage at longer wavelengths would maximize the science outcome of all planned observatories.

Three decades after its launch, Hubble still has not reached its full potential. Hubble remains a powerful tool for space exploration and can continue to do so for many years provided we preserve and enhance its functionality.  The latest Hubble reports\footnote{http://www.stsci.edu/institute/stuc/presentations} provide a clear picture of the status of different parts of the telescope including its science instruments, and its pointing control system. {\bf All the science instruments on Hubble are operating nominally, and engineering analyses predict about 80\% probability of science operations continuing beyond 2025\footnote{http://www.stsci.edu/institute/stuc/may-2019/Hubble2025.pdf}. The highest risk limitation for Hubble to continue operations at this point is the pointing control system}. 

Hubble's pointing control system has a total of six gyroscopes, all replaced during SM4 in 2009. The electronic leads on these gyroscopes degrade over time, and only three of the six remain operational. In October 5, 2018 Hubble went into {\it safe mode} for over two weeks after the last spare gyroscope failed. The Hubble operations team managed to bring the observatory back to regular operations with the last three working gyroscopes, all of which feature so-called enhanced flex leads designed to be more robust for longer operational life. Using these, Hubble will be able to run in three-gyroscope mode until another gyro fails (currently expected to be in the early 2020s). At that point the telescope will switch to  one-gyroscope mode, with the other remaining gyro switched off in reserve for later use. While Hubble's science capabilities will remain high in one-gyro mode, pointing restrictions will reduce science efficiency. In this scenario eventually the telescope will have all science instruments still functioning, but will no longer be able to point. The prediction is that this may happen around the mid-2020s.

{\bf In this White Paper we encourage NASA and ESA to explore the technical feasibility and risk of a sixth servicing mission to Hubble to refresh its gyroscope set and update other telescope parts to prolong Hubble's operations past the 2020s, including, if feasible, the installation of new UV-enabled imaging and spectroscopic instruments}. In the next section we briefly highlight examples of key science cases that extending the lifetime of Hubble would enable. We then describe recent advances in robotic servicing that establish the feasibility of fully robotic servicing as the next step in keeping Hubble at astronomy's forefront for many years to come.

\vspace{-0.5cm}
\begin{table}[ht!]
\setlength{\arrayrulewidth}{0.5mm}
\setlength{\tabcolsep}{12pt}
\renewcommand{\arraystretch}{1.4}
\colorlet{lteal}{teal!25!}
\colorlet{lcyan}{cyan!25!}
\begin{tabular}{|p{1.6cm}|p{2.88cm}|p{2.0cm}|p{5.0cm}|} 
\hline
\small \cellcolor{lightgray} Servicing Mission & \cellcolor{lightgray} Dates & \cellcolor{lightgray} New Instruments & \cellcolor{lightgray} Other Components Installed\\
\hline\hline
\small \cellcolor{lteal} SM1 & \cellcolor{lteal} Dec 2-3, 1993 & \cellcolor{lteal} COSTAR* & \cellcolor{lteal} Solar Arrays  \\
\small \cellcolor{lteal}    &  \cellcolor{lteal}             & \cellcolor{lteal} WFPC2   & \cellcolor{lteal} Magnetometers\\
\small \cellcolor{lteal}    &   \cellcolor{lteal}            & \cellcolor{lteal} & \cellcolor{lteal}Coprocessors\\
\small \cellcolor{lteal}    &   \cellcolor{lteal}            & \cellcolor{lteal} & \cellcolor{lteal}Gyroscopes\\
\small \cellcolor{lteal}    &    \cellcolor{lteal}           &  \cellcolor{lteal}& \cellcolor{lteal}GHRS Redundancy Kit\\
\hline
\small \cellcolor{lcyan} SM2 & \cellcolor{lcyan} Feb 11-21, 1997   & \cellcolor{lcyan} STIS & \cellcolor{lcyan} Refurbished FGS\\
\small \cellcolor{lcyan}    &      \cellcolor{lcyan}             & \cellcolor{lcyan} NICMOS & \cellcolor{lcyan} Optical Control Electronics Enhancement Kit\\
\small \cellcolor{lcyan}    &      \cellcolor{lcyan}             & \cellcolor{lcyan} & \cellcolor{lcyan} Solid State Recorder \\
\small \cellcolor{lcyan}    &     \cellcolor{lcyan}              & \cellcolor{lcyan}  & \cellcolor{lcyan} Reaction Wheel Assemblies \\
\small \cellcolor{lcyan}    &    \cellcolor{lcyan}               & \cellcolor{lcyan}  & \cellcolor{lcyan} Data Interface Units\\
\small \cellcolor{lcyan}    &    \cellcolor{lcyan}               &  \cellcolor{lcyan} & \cellcolor{lcyan} Solar Array Drive Electronics\\
\hline
\small \cellcolor{lteal} SM3A & \cellcolor{lteal} Dec 19-27, 1999 & \cellcolor{lteal} --- &  \cellcolor{lteal}  Gyroscopes\\
\hline
\small \cellcolor{lcyan} SM3B & \cellcolor{lcyan} Mar 1-12, 2002 & \cellcolor{lcyan} ACS & \cellcolor{lcyan} Solar Array 3\\
\small \cellcolor{lcyan}     &  \cellcolor{lcyan}              &  \cellcolor{lcyan}   &\cellcolor{lcyan}  Power Control Unit \\
\small \cellcolor{lcyan}     &      \cellcolor{lcyan}          & \cellcolor{lcyan}    & \cellcolor{lcyan} NICMOS Cryocooler \\
\hline
\small \cellcolor{lteal} SM4 & \cellcolor{lteal} May 11-24, 2009 & \cellcolor{lteal} COS & \cellcolor{lteal} STIS \& ACS fixed\\
\small \cellcolor{lteal}     &    \cellcolor{lteal}            & \cellcolor{lteal} WFC3 & \cellcolor{lteal} Replace Gyroscopes\\
\small \cellcolor{lteal}     &    \cellcolor{lteal}            &   \cellcolor{lteal}   & \cellcolor{lteal} Replace Batteries\\
\small \cellcolor{lteal}     &     \cellcolor{lteal}           &   \cellcolor{lteal}   & \cellcolor{lteal} Refurbished FGS\\
\hline

\end{tabular}
\caption{ Dates, new instruments installed, and other components installed during each of the five Hubble service missions between 1993 and 2009.}
\end{table}

\vspace{-1.0cm}
\begin{figure*}[h!]
\centering
\includegraphics[angle=270, scale=0.5, trim= 0cm 0cm 2.5cm 0cm, clip=true]{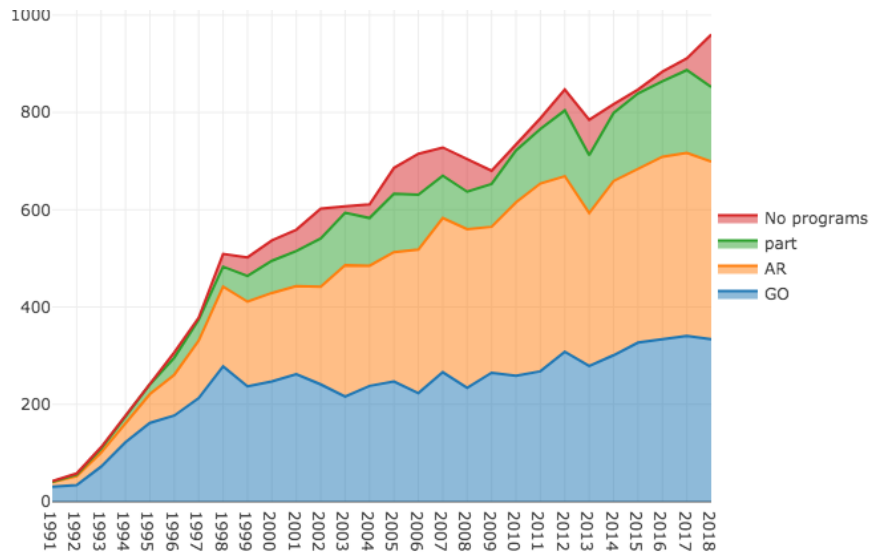}
\vspace{-0.2in}
\caption{Number of refereed papers per year between 1991 and 2018 that used data from Hubble. ${\it GO}$ and ${\it AR}$ refer to General Observer and Archival programs. ${\it Part}$ refers to publications that use data from both GO and AR programs. {\it No programs} refers to publications that don't explicitly identify the Hubble program their data came from. \label{fig:publications}}
\end{figure*}

\vspace{2mm}
\noindent {\bf Key Science Goals and Objectives}
\vspace{1mm}



As the next generation of powerful large telescopes on the ground (GMT, TMT and the European ELT), and space (JWST, WFIRST) come online over the next decade, Hubble will still remain the best facility for providing high-resolution, deep UV and optical imaging, and UV spectroscopy. Preserving (or ideally augmenting Hubble's capabilities by installing improved instrumentation on it), will be critical to maximize the scientific return from the next generation of telescopes. Examples of science that will be enabled using Hubble's UV/blue optical capabilities include:

{\bf - Solar System Planet:} Hubble has been key to begin characterizing many Solar System objects, given the many key atomic and molecular spectral signatures in the UV. 
For example, Hubble revealed very diverse auroral emissions on Jupiter, Saturn and Uranus (e.g., Grodent 2015; Clarke et al.\ 1996,2002; Hill 2001; Ballester 1996; Lamy et al.\ 2012). Auroral emissions reflect the state and processes of a magnetosphere that can be related to both internal and external-solar-wind inputs, but there is still for example controversy as to whether the Jovian magnetosphere is dominated by its internal sources or not (Grodent et al.\ 2018). The JUNO spacecraft is currently sampling different regions of Jupiter's magnetosphere. JUNO's data will be interpreted together with Hubble data, as done with Saturn during the Cassini Grand Finale.
Hubble has also revealed the atmospheric composition of the Galilean satellites, and their unique interactions with the Jovian plasma. Io's atmosphere has shown volcanic and frost-sublimation sources, and geometrically complex signatures of 
plasma interactions 
(Ballester et al.\ 1994; Roesler et al.\ 1999). Europa has shown water plumes originating from active processes in its sub-surface ocean, 
and variable global emissions different to those on Io 
(Roth et al.\ 2014).  Ganymede has shown 
unique auroral ovals, 
creating a well defined “mini” magnetosphere within the immense Jovian magnetosphere (Feldman et al.\ 2000; Paty \& Winglee 2004). Hubble's global, spatial and temporal context would be key for the upcoming JUICE (Jupiter Icy Moons Explorer) mission and also for the Europa Clipper in the late 2020's to continue studying all these solar system bodies.
Hubble has also revealed extended H\,I and the D/H ratio over long periods in Mars upper atmosphere complementing the MAVEN mission (Clarke et al. in progress), and the distribution of SO2 in Venus' atmosphere supporting the Venus Express mission (Jessup et al.\ 2015). For Saturn, Hubble is providing details of the H\,I in the upper atmosphere essential to interpret Cassini UV data (Ben-Jaffel et al. in progress).

{\bf - Characterization of exoplanetary atmospheres}: Some 
key information pieces 
to understand the physics and chemistry of exoplanetary atmospheres are their cloud coverage, 
atmospheric escape rates, and the amount of high energy stellar irradiation 
the planet endures. 
Some atmospheric escape can be measured from the ground in the near-infrared (e.g., Helium; Spake et al.\ 2018), but most of this information can only be obtained from Hubble's observations in the UV and blue optical 
Without UV and blue optical observations (especially those yielding information about the atmospheric heights of clouds), chemical composition and mixing ratios of the main absorbers in exoplanetary atmospheres cannot be determined, even if they show robust absorption features in the infrared (see e.g.\ Benneke \& Seager 2012, and Fig.~4 in Sing et al.\ 2016).  
The properties of the uppermost planetary layers and their escape are key to understanding and modeling the atmospheric physics of exoplanets, and most everything currently known regarding these uppermost layers is based on Hubble UV
transit observations including H\,I in Ly$\alpha$, 
O\,I, C\,II, and Mg\,I. Atmospheric escape can be substantial, and
significantly affect the evolution of low-mass planets (e.g., Lecavelier 2007; Ehrenreich et al.\ 2015; Owen \& Alvarez
2016), as the bulk of the primordial accreted volatile atmosphere can be eroded away by atmospheric escape. Finally, measurements of the UV emission from the host star are key to characterizing the atmospheric properties of exoplanets, especially small planets in the habitable zones of nearby stars discovered by the NASA TESS mission (Ricker et al.\ 2016). Those planets will be the main targets for future bio-signature searches, but without Hubble, measurements of the UV stellar irradiation levels on those planets will be delayed until a mission like LUVOIR\footnote{https://asd.gsfc.nasa.gov/luvoir/} or HabEx\footnote{https://www.jpl.nasa.gov/habex/} launches in the 2030-40s.

\indent {\bf - Understanding how exoplanets assemble their atmospheres}: A critical next step in exoplanet studies is understanding how the planets assemble and grow their nascent atmospheres.  Planet-forming disks provide 
context for spectroscopy of gaseous and super-Earth planets (with JWST), and lay the groundwork for future studies of potential Earth-like planets (in the 2030-40s with LUVOIR life-finding missions; e.g. France et al.\ 2019).  UV spectroscopy 
enables observations of molecular gas in the inner regions of protoplanetary disks 
and mass accretion onto their protostellar hosts (Ingleby et al.\ 2013; Ardila et al.\ 2013; France et al.\ 2012).  UV fluorescent H$_{2}$ spectra are sensitive to gas surface densities lower than 10$^{-6}$ g cm$^{-2}$,  making them an extremely useful probe of remnant gas at r $<$ 10 AU (Ingleby et al.\ 2012). UV absorption line spectroscopy provides direct access to key molecular species such as H$_{2}$, CO, OH, H$_{2}$O.  This is the only direct observational technique able to characterize co-spatial populations of molecules with H$_{2}$, offering unique access to absolute abundance and temperature measurements (Roberge et al.\ 2001; France et al.\ 2014). Installation of a high-sensitivity, multi-object UV spectrograph 
would enable high-efficiency surveys of disk systems in star-forming regions with high protostar density (e.g., the Orion Nebula Cloud), completing the census of protoplanetary disk lifetimes, radial  structures, and abundances started with existing Hubble programs and the upcoming ULLYSES project.

{\bf - Understanding the origin of multiple stellar populations in globular clusters (GCs) and unveiling their internal kinematics:} 
GCs were thought to be made of stars with the same age and chemical composition as a result of a single star formation episode. The groundbreaking discovery that virtually all relatively massive stellar clusters host multiple stellar populations with different light-element abundances (Gratton et al.\ 2012; Piotto et al.\ 2015; Bastian \& Lardo 2018) has de-facto opened a new era in stellar astrophysics research. Hubble observations were key to this discovery, by providing the key combination of UV and optical colors (Milone et al.\ 2017). 
The critical next challenge is understanding the 
physical mechanisms shaping the multiple-population properties, something that can be only done with Hubble's repowered UV capabilities. 
GCs are also the only astrophysical systems that within a Hubble time, undergo nearly all the physical processes known in stellar dynamics. 
This is why the systematic measure of high-quality proper motions (PMs) for individual stars in Galactic GCs was vigorously promoted and carried out with Hubble in recent years (Bellini et al.\ 2014,2017). The stable environment of space makes Hubble an unprecedented astrometric tool, surpassing Gaia in densest GC regions. 
 Hubble also is the only instrument allowing accurate PM measures of stars along the entire main sequence 
Accurately measuring individual star PMs in Galactic GCs  will yield 
a quantum jump in our comprehension of multi-body dynamics and to firmly address the existence of 
intermediate-mass black holes in the center of GCs, considered to be invaluable sources of gravitational wave emission. 

{\bf - Exotic Stellar Populations in Star Clusters:}
The high-density, central regions of clusters are inaccessible with ground-based telescopes. They are key regions to understand the formation of collisionally induced stellar populations. Close gravitational interactions between stars alter the overall energy budget and can affect stellar evolution, generating exotic objects like blue straggler stars, millisecond pulsars, X-ray binaries, and cataclysmic variables (e.g., Bailyn 1995). Thanks to its unprecedented combination of high-spatial resolution and UV capabilities, 
Hubble gave birth to the science of stellar exotica, opening unexplored astrophysical research.
We can now use blue stragglers as an empirical clock to measure the level of dynamical evolution of stellar clusters (Ferraro et al.\ 2012, 2018). Moreover, the systematic search for the optical companions to binary millisecond pulsars (Ferraro et al.\ 2001,2015; Pallanca et al.\ 2010; Cadelano et al.\ 2015), is starting to draw a coherent evolutionary path in the scenario that connects X-ray binaries to these exotica. 
Progress on our understanding of collisional-induced stellar populations and the interplay between dynamics and stellar evolution will be halted for decades without UV capabilities like Hubble’s.

{\bf -Understanding star formation processes:}
Hubble UV observations of star forming galaxies in the Local Universe have provided crucial information on their recent ($<$50\,Myr) star formation histories from resolved massive stars, and the direct measures of ages and masses of star clusters and associations (e.g., LEGUS; Calzetti et al.\ 2015). However, many aspects still need to be understood, and the extension of these observations to the full range of morphologies, star formation rates, masses, metallicities, internal structures, and interaction states is key for our understanding of the clustering and star formation processes, and the impact of the local environment on them. A re-powered Hubble UV capability would enable robust characterization of these processes
to understand the relation between gas content and state, and star formation, and of the true nature of the clumpy star formation patterns observed at high redshift.

{\bf -Interstellar Medium:} The local interstellar medium (LISM) -- the gas and dust that surrounds the Sun and the nearest stars -- is only accessible via high resolution UV spectroscopy (Frisch et al.\ 2011). 
Not only does the LISM provide the closest example of general interstellar medium phenomena that occur throughout our galaxy and beyond (e.g., star formation, Evans 1999; and supernovae McCray \& Snow 1979), but is also intimately tied to the interaction of stellar winds and the surrounding LISM. The heliosphere is defined by the pressure balance between the solar wind and the LISM (McComas et al.\ 2012), and analogous structures (astrospheres) have been detected around other nearby stars (Wood et al.\ 2005). These interactions have profound impacts on the structure of the interplanetary medium and control the cosmic ray flux incident on the planetary atmospheres orbiting these stars (Zank \& Frisch 1999). In the era of evaluating the habitability of exoplanets, such considerations will be important and access to the UV vital. 

{\bf -Degenerate Accreting Systems:} The UV is essential for studying the physics of cataclysmic variables and related degenerate accreting systems.  For accretion disk physics, the UV samples from the boundary layer of neutron stars and white dwarfs to the interaction region between the periphery of the disk and the inflowing matter from the companion.  Symbiotics, which also harbor degenerates but accrete from winds of giant companions, also display most of their essential phenomenology in the UV.  Both recurrent and classical novae cannot be effectively understood unless well observed in UV, as their optical and longer wavelength behavior is determined by opacity changes in the mid-UV during the expansion.
Other than Swift grism spectroscopy
(longward of 1700\AA\ and missing almost all important resonance lines of high metallic ions), no facility offers high spectroscopic resolution.  These transient phenomena each display unique spectral development, but governed by common underlying physical mechanisms that still remain to be disentangled.  Unless the UV is available, understanding this entire population will be severely impeded.  

{\bf -Better understanding of supernovae (SNe) progenitors:}
Hubble is presently investing significant time 
on the progenitors of SNe of thermonuclear, core collapse, magnetar or 
pulsational instability  origin.  Direct detections of core collapse SNe progenitors have been 
possible via high resolution Hubble images. 
The study of individual massive stars in resolved galaxies out 
to $\sim$20 Mpc has become common, as well as successful identification of pre-explosion sources (see Smartt 2009; Van Dyk et al.\ 2003, 2019).  With an extended Hubble mission these studies can continue into the JWST era, where supernovae from massive stars are more easily identified at UV and blue optical wavelengths. No direct detection of a progenitor of a Type Ia supernova has yet been made using this method. However, the impact on the companion 
star of the white dwarf that explodes as a SN Ia can be studied in various 
ways. The PM measured by Hubble 
of a surviving star has to be significant in 
comparison to the rest of the stars within the innermost part of the 
remnant (see Bedin et al.\ 2014; Ruiz-Lapuente 2018 for a review),
 and deep imaging of the remnants 
in the LMC can exclude or confirm the presence of a companion star 
(see, e.g., Schaefer \& Pagnotta 2012, Li et al.\ 2017). Another approach to the progenitor problem is through the shock breakout 
of the supernova ejecta when they impact the circumstellar medium or 
the companion (see, e.g., Graham et al.\ 2019; Bersten et al.\ 2018;
 Bersten et al.\ 2013).
 For all types of SNe, a flash is expected in the UV,
and obtaining UV spectra at this very early stage is key to 
identify the nature of the explosion. Ultimately, the understanding of SNe Ia through UV spectra will 
reaffirm their cosmological use.

{\bf - Escape fraction of ionizing photons:} Star-forming galaxies likely dominated the ionizing photon budget for the reionization of the intergalactic medium, but little is known as to how ionizing photons escape the interstellar and circumgalactic media of their host galaxies.  Significant progress has been made, led by deep Hubble/COS spectroscopy, to directly detect escaping ionizing photons from low-redshift star-forming galaxies
(e.g., Puschnig et al.\ 2017).  To understand the implications of these observations for high-redshift galaxies, we need to observe large numbers of galaxies, spanning a wide range in stellar mass, star-formation rate, dust attenuation, ISM geometry, etc., 
While some such programs are underway now
with COS, 
we will not reach the sample sizes needed before the end of the Hubble mission. The same remains true for attempts to directly detect escaping ionizing photons from galaxies at redshifts 2\,$<$\,$z$\,$<$\,4 through analyses of stacked images at rest-frame far-UV (observed mid- and near-UV) wavelengths.  The problem of contamination by interlopers at lower redshifts, but very near the line-of-sight, can only be solved at the high resolution offered by WFC3/UVIS on Hubble (e.g., Siana et al.\ 2010; Smith et al.\ 2018).

{\bf - Studying individual stars at cosmological distances:} Hubble discovered the first star observed at cosmological distances (at $z=1.49$), dubbed Icarus, thanks to a microlens near the critical curve of a gravitational lens (the galaxy cluster MACS J1149) (Kelly et al.\ 2018). The combined magnification from the cluster and the microlens increased the observed flux of the distant background  star by a factor of more than 2000. Such fortuitous alignments, although rare, are expected to happen more often near the critical curves of large gravitational lenses. More candidates have been identified already (Rodney et al. 2018, Chen et al. 2019). Hubble observations can provide the stars' spectral types, their temperature and age. A larger number of stars will allow to understand the formation and evolution of massive stars at $z>0.5$. Hubble's sensitivity in the UV and optical bands is critical to characterizing the properties of the lensed stars, since many luminous stars have hot photospheric temperatures and the Balmer break lies in the optical for lensed stars with redshifts of up to approximately one.

{\bf -Cosmology in the Local Group:} There is growing evidence that satellite galaxies are aligned in flattened planes around the Milky Way, Andromeda, and other more distant, massive galaxies.  However, such phase-space distributions are rare in cosmological simulations based on the $\Lambda$ CDM paradigm, and it has become increasingly clear that the satellite system around the Milky Way is not typical in terms of its stellar population, quenching properties, or internal and external kinematics.  The Andromeda satellite system is therefore critical for testing the stability of galaxy planes and hence the CDM paradigm (e.g., Boylan-Kolchin et al.\ 2012; Simon 2019), the growth of galaxies over time and the relationship between dwarf galaxies and reionization (e.g., Brown et al.\ 2014, Weisz et al.\ 2014).  PM measurements, only possible currently with Hubble, are needed to assess whether the ‘Great Plane’ of satellites is dynamically stable and rotating.  Star formation in very low mass galaxies is expected to be suppressed or quenched completely by reionization (e.g., Bullock et al.\ 2000; Ricotta \& Gnedin 2005).  The timing and length of this suppression, determined from deep color-magnitude diagrams that reach below the main sequence turnoff (e.g., Brown et al.\ 2014), gives important insight into the main sources of reionization.  The inner mass profiles within dwarf galaxies have emerged as the most important test of the nature of dark matter, motivated by the ``core-cusp'' and ``Too Big to Fail'' problems (e.g., Boylan-Kolchin et al.\ 2011). However, PMs of individual stars within dwarf galaxies are needed to break projection/orbital degeneracies (Evslin 2015), thus enabling robust measurements of their 3-D mass profiles (Strigari et al.\ 2007). The most stringent tests come from the Ultra-faint galaxy population, many of whose stars are too faint for Gaia, and require Hubble and stability over long time baselines.

{\bf -Probing the physics of galactogenesis:} How galaxies assemble and how the Hubble sequence emerges remain open questions in astrophysics and cosmology.  They are among the pillars of NASA's COSMIC ORIGINS program. 
Hubble has played a key role in exploring the evolution of galaxies from
a few hundred million years after the Big Bang to the present (Finkelstein 2016; Bouwens 2018). The discovery of galaxies up to redshift z\,$\sim$\,11 (Oesch et al.\ 2018), reconstruction of the assembly of the stellar body of massive galaxies (Barro et al.\ 2017), the quenching of star-formation (Whitaker et al.\ 2017; Fang et al.\ 2018) and the morphological and structural transformations that lead to the Hubble sequence (Guo et al.\ 2017), the interplay between stars, AGN (Kocevski et al.\ 2017), the ISM, and the gas in the Circum Galactic Medium (Tumlinson et al.\ 2017),  
are all still in their infancy, yet Hubble remains the \textbf{\itshape only} observatory capable of continuing this research until the advent of LUVOIR.  
Hubble has spurred
great progress in the analysis of panchromatic data sets. Techniques such as
photometric redshift, SED fitting and star-formation history reconstruction (Leja et al. 2019) are
powerful tools 
given panchromatic and well sampled data 
from the rest-frame UV to
the optical. The lack of Hubble data will be the
limiting factor in these studies (e.g., Guo et al.\ 2018).
Expanding the lifetime of Hubble
with new, improved photometric and spectroscopic capabilities will allow us continue the empirical exploration of all these topics. Only Hubble can
provide the UV-visible sensitivity and angular resolution to complement JWST and thus enable panchromatic studies.

{\bf - Counterparts of Gravitational Wave emitters:} The Swift satellite detected surprisingly intense UV emission in the early phases of 
the neutron star merging event GW170817 (Evans et al.\ 2017), revealing that UV observations are key to 
characterize the merging process of degenerate star systems. 
In the forthcoming LISA era, we should be able to detect GW signals even from double white dwarf mergers. These systems (and their precursors) are expected to stand out in the UV, especially in globular clusters that are the richest nearby reservoirs of such objects. The refurbished UV capability of Hubble is the only chance we have to properly exploit 
the gold mine of information brought by the UV radiation from the electromagnetic counterparts of GW emitters.
Both, double white dwarfs (WDs), and white dwarfs in binaries with black holes or neutron stars are expected 
source of gravitational radiation with LISA.  
Two of three GCs ultracompact X-ray binaries have had their orbital periods measured in the UV (Dieball et al.\ 2005; Zurek et al.\ 2009), and the best candidate for a GC WD-WD binary has also come from UV variability studies (Zurek et al.\ 2016).  The surface has only been scratched for what Hubble UV variability studies of GCs can do, but 
if done before the LISA launch, this work will have the potential for allowing more sensitive directed searches for GWs from these objects.  If the Hubble UV mission extends well into the LISA era, it may also allow for follow-up of LISA sources both in clusters and in the field.

{\bf - UV/optical imaging to complement JWST:} While JWST will enable deep near-infrared imaging of distant galaxies, its throughput drops dramatically at $\lambda$\,$<$\,0.9 $\mu$m, and it is incapable of probing $<$ 0.6 $\mu$m.  Deep 0.3--0.7 $\mu$m imaging over the same regions that JWST will probe provides critical wavelengths for Lyman break dropout studies at $z$\,$>$\,5, as well as the rest-UV for galaxies at $z$\,$\sim$\,1--5.  While some imaging is in place (e.g., UVUDF) or in progress (e.g., UVCANDELS, JWST NEP Time-Domain Field), unless Hubble is serviced, it will be impossible to obtain such observations in the future for new JWST survey regions, or as follow-up of JWST discoveries.


\vspace{3mm}
\noindent {\bf Technical Overview}
\vspace{1mm}

The key need for Hubble to continue operations is refreshing the gyroscopes used in its pointing control system. The gyroscopes on Hubble have already been replaced three times -- during SM1, SM3A and SM4 -- as shown in Table 1. The last set of gyros (installed in 2009) has extended the lifetime of Hubble for at least 10 years. {\bf A new servicing mission could install a new set of enhanced flex lead gyroscopes and extend the productive lifetime of Hubble into at least the 2030s}. Other systems such as computers and batteries appear currently to be in relatively good health, but as part of the servicing mission development NASA would naturally assess whether these or other components should be refreshed to maximize mission lifetime.

The second greatest need is in UV instrumentation, particularly given the finite lifetime of COS.
On the instrumentation side, we (Hubble users signing this white paper) think that installing a new multi-object spectrograph on Hubble would greatly enhance its scientific productivity. A UV multi-object spectrograph (e.g., France et al.\ 2017), building on advances in UV detector technology (Siegmund et al.\ 2017; Nikzad et al.\ 2017), optical coatings (Fleming et al.\ 2017), and JWST microshutter array work and subsequent development for LUVOIR and Habex (McCandliss et al.\ 2019), would be transformative for Hubble.  Such an instrument should have resolving power of $>$20,000 from 100--350 nm, cover a field-of-view larger than (2$'$\,$\times$\,2$'$), and have FUV effective area the same or higher than COS and NUV effective area 3$\times$ that of Hubble-COS.

\textbf{The time is right for a robotic servicing mission to Hubble.} The possibility of Robotic servicing of Hubble was studied in depth in the period pre SM4 (e.g.\ Oda et al.\ 2005, Wang et al.\ 2006). 
Substantial advances in space robotics over the past decade have made robotic servicing a truly viable option today (Gefke et al. 2015). 
Robotic servicing is now being widely pursued for commercial satellite operations in low earth orbits (LEO) and especially geostationary orbits (GEO). Capabilities for in-space robotic servicing have been progressively proven by a series of technology demonstrator missions, and are leading towards operational missions in the immediate future. The Defense Advanced Research Projects Agency (DARPA) Orbital Express flew two spacecraft to demonstrate autonomous rendezvous, capture and berthing, on-orbit refueling, and even robotic replacement and upgrade of electronics modules (Friend 2008). All objectives were successfully achieved during a two month mission in 2007, including robotic transfer and installation of battery modules and a computer upgrade module, in some ways comparable to replacing an RSU (gyroscope) module on Hubble. Many servicing techniques first pioneered for Hubble were later employed and improved on during the assembly of the International Space Station. NASA now has decades of expertise in flight performing highly sophisticated assembly and maintenance tasks via tele-operation from the ground of the station's robot arms. Starting in 2011, the Robotic Refueling Mission has demonstrated at the ISS a wide range of remote servicing techniques for refueling and repairing satellites. Many of the robotic tools used in RRM have direct technology heritage to the custom tooling developed for astronaut use in servicing Hubble. 

These technology demonstrations are now leading to real flight robotic servicing projects 
NASA's next step is the RESTORE-L mission (Reed et al. 2016), an upcoming flight project that will demonstrate robotic refueling and servicing of the Landsat-7 spacecraft. RESTORE-L is scheduled for launch in late 2021 or early 2022. 
Meanwhile, other US and European industry efforts are also advancing space servicing capabilities for commercial and military customers. In September 2019, Northrop Grumman will launch their first Mission Extension Vehicle (MEV), on a commercial contract to dock with and extend the lifetime of an Intelsat communications satellite. A second MEV is scheduled to launch in 2020. 2020 should also see the first launch of two ``Space Drone'' servicing spacecraft from the UK company Effective Space. Other firms such as SSL/Maxar, Thales Alenia, are also advancing relevant capabilities. Commercial space servicing is quietly becoming the new reality.

Taking advantage of these capabilities it is possible
to envision robotic servicing of Hubble. NASA already has deep expertise in planning and executing intricate tasks on and around Hubble, and all interfaces are well understood. SM4 left Hubble equipped with a Soft Capture Mechanism including a low-impact docking adapter and navigation targets to assist autonomous proximity operations and docking. A robotic spacecraft, likely modeled on that of RESTORE-L, would rendesvous and dock with Hubble, and then would use robotic manipulators to open access panels and install new modules or instrumentation as desired. Some tasks such as docking would likely happen under local autonomy, while servicing tasks would likely be methodically teleoperated from the ground, using tried-and-true methods from ISS operations.

\vspace{3mm}
\noindent {\bf Technology Drivers}
\vspace{1mm}

The necessary technologies for remote servicing are already being pursued 
by NASA's Satellite Servicing Projects Division and by commercial firms, leading to flight demonstration missions in a few years as described above.  
Replacement gyroscope modules could 
be crafted as copies of the existing enhanced flex lead gyros; a trade study could also evaluate whether newer technologies such as hemispherical resonator gyros could 
be interfaced to Hubble's legacy systems.

{\bf Achieving a next servicing mission for Hubble would not only extend its scientific lifetime by years, it could set the stage for future servicing of NASA's next flagships}. WFIRST is being designed for robotic servicing and refueling, including modular interfaces with heritage to Hubble. Plausibly, with servicing WFIRST could be operational for \textit{decades}, making it a worthy successor to Hubble in longevity as well as scientific capabilities.  While 
not well suited for servicing in most regards, 
JWST could be refueled potentially to extend its lifetime too. Furthermore, the same technologies and capabilities for servicing are also applicable to in-space assembly of large observatories, which recent NASA study suggests may reduce risk and potentially lower costs for future missions, as well as
opening up long-term paths to observatories beyond the limits of any single launch vehicle. See white paper by Mukherjee et al.

\vspace{3mm}
\noindent{\bf Organization, Partnerships, and Current Status:}
\vspace{1mm}

This white paper seeks to encourage NASA, ESA, the science community, and the space technology private sector to assess the scientific merit, technical feasibility, and risks of another servicing mission to Hubble, to replace gyroscopes and enhance its instrumentation. {\bf Given Hubble's unique capabilities, there is a desire by a significant fraction of the astrophysics community to look into another possible servicing mission to Hubble}, manned or robotic. While there is not yet a specific organization or partnerships set for this proposal, NASA's and industry's growing capabilities in space servicing appear directly suited to this task. 
Though we cannot yet present a detailed cost estimate for such a mission, we can use NASA's RESTORE-L mission as a baseline. 
The RESTORE-L demonstration mission is expected to cost around \$800M. As this is a first-of-its-kind technology demonstration mission, some of that cost has been driven by developing and qualifying new capabilities. Such costs might be reduced on a future mission to Hubble reusing similar systems as RESTORE-L. Additional funds would be needed for development of new instrumentation, likely of order a few hundred million dollars. Thus it may be reasonable to estimate the cost of SM5 to Hubble at approximately \$800M-\$1B, in addition to extended mission telescope operating costs. {\bf For roughly the cost of 1-2 probe class missions, we can have another decade of flagship ``Great Observatory scale'' Hubble science - while also setting the stage for sustainable, serviced, long-lived Great Observatories across many future decades}.





\pagebreak
\noindent \textbf{References}

Ardila, D.; Herczeg, G.; Gregory, S. et al., 2013, ApJS, 207, 1

Bailyn, C., 1995, ARA\&A 33,133

Ballester, G.E. et al. 1994, Icarus, 111, 2

Ballester, G.E. et al. 1996, Science, 274, 409

Barro, G., Faber, S.M., Koo, D.C., et al., 2017, ApJ, 840, 47   

Bastian, N. \& Lardo, C. 2018, ARA\&A, 56, 83

Bedin, L., Ruiz-Lapuente, Gonz\'alez Hern\'andez, J.I., et al. 2014, MNRAS, 
439, 354

Bellini, A.; Bianchini, P.; Varri, A. et al. 2017, ApJ, 844, 164

Bellini, A.; Anderson, J.; van der Marel, R. et al., 2014, ApJ, 797, 115 

Benneke, B. \& Seager, S. 2012, ApJ, 753, 100

Bersten, M.C. Folatelli, G. Garcia, F et al. 2018, Nature, 554, 497 

Bersten, M.C., Tanaka, M, Tominaga, N. 2013, ApJ 797, 143

Boylan-Kolchin, M.; Bullock, J. S.; Kaplinghat, M. et al., 2012 MNRAS, 422, 1203

Bouwens, R., 2018, Nature, 557, 312  

Brown, T. M.; Tumlinson, J.; Geha, M. et al., 2014 ApJ, 796, 91

Bullock, J. S.; Kravtsov, A. V.; Weinberg, D. H. et al., 2000 ApJ, 539, 517

Cadelano, M., Pallanca, C., Ferraro, F.R., et al., 2015, ApJ, 812, 63

Calzetti, D., Lee, J.C., Sabbi, E., et al., 2015, AJ, 149, 51

Chen, W., Kelly, P.L., Diego, J.M., et al., 2019, arXiv:1902.05510

Clarke, J.T. et al. 1996, Science, 274, 404

Clarke, J.T. et al. 2002, Science, Nature, 415, 997

Dieball et al. 2005 ApJ 634L 105

Ehrenreich, D., Bourrier, V.; Wheatley, P. et al., 2015, Nature, 522, 459

Fang, J.~J., Faber, S.M., Koo, D.C., et al., 2018, ApJ, 858, 100  

Feldman, P.D. et al. 2000, ApJ, 535, 1085

Ferraro, F.R., D'Amico, N., Possenti, A., Mignani, R.P., \& Paltrinieri, B., 2001, ApJ, 561, 337 

Ferraro, F.R., Lanzoni, B.; Dalessandro, E. et al., 2012, Nature, 492, 393

Ferraro, F.R., Pallanca, C., Lanzoni, B., et al., 2015, ApJ, 807, L1

Ferraro, F.R., Lanzoni, B., Raso, S., et al., 2018, ApJ, 860, 36 

Finkelstein, S.L., 2016, PASA, 33, 37

Fleming, B.; Quijada, M.; Hennessy, J. et al., 2017, Appt, 56, 9941

France, K.; Pascucci, I.; Dong, R. et al., 2019, BAAS, 51, 167

France, K.; Fleming, B.; West, G. et al., 2017, SPIE, 10397, 13

France, K.; Herczeg, G.; McJunkin, M. et al. 2014, ApJ, 794, 160

France, K., Schindhelm, E.; Herczeg, G. et al., 2012, ApJ, 756, 171

Friend, R. et al.; 2008, Proc SPIE. 6958

Graham, M.L, Harris, C.E., Nugent P.E et al. 2019, 871, 62

Gefke et al. 2015, Proc. AIAA Space 2015.

Gratton, Carretta, E. \& Bragaglia, A., 2012, ARA\&A, 20, 50 

Grodent, D. 2015, Sp. Sci. Revws., 187, 23

Grodent, D. et al. 2018, J. Geophys. Res. Space Physics, 123, 3389

Guo, Y., Bell, E.F., Lu, Y., et al., 2017, ApJL, 841, L22

Guo, Y., Rafelski, M., Bell, E.F., et al. 2018, ApJ, 853, 108  

Hill, T.W. 2001, J. Geophys. Res., 106, 8101

Ingleby, L.; Calvet, N.; Herczeg, G. et al. 2013, ApJ, 767, 112

Ingleby, L.; Calvet, N.; Herczeg, G. et al., 2012, ApJ, 752, 20

Kelly, P.L., Diego, J.M., Rodney, S. et al., 2018, Nature Astronomy, 2, 334.

Kocevski, D.D., Barro, G., Faber, S.M., et al., 2017, ApJ, 846, 112

Lamy, L., Prangé, R., Hansen, K. C., et al., 2012, Geophys. Res. Let., 2012, 39, L07105.

Lecavelier Des Etangs, A., 2007, A\&A, 461, 1185L

Leja, J., Carnall, A.C., Johnson, B.D., Conroy, C., \& Speagle, J.~S., 2019, ApJ, 876, 3  

Li, Ch.-J., Chu, Y-H, Gruendl, R.A., et al. 2017, ApJ, 839, 85

McCandliss, S.; Calzetti, D.; Ferguson, H. et al., 2019, BAAS, 51, 535

Milone, A.; Piotto, G.; Renzini, A. et al., 2017, MNRAS, 464, 3636

Nikzad, S.; Jewell, A.; Hoenk, M. et al.  2017, JATIS, 3, 036002

Oda, M. et al.\ 2005, Proceedings of ISAIRAS conference 2005, 603, 9

Oesch, P.A., Bouwens, R.~J., Illingworth, G.D., Labb{\'e}, I., \& Stefanon, M., 2018, ApJ, 855, 105  

Owen, J. \& Alvarez, M., 2016, ApJ, 816, 34O

Ozel, F. \& Freire, P. 2016, ARA\&A, 54, 501

Pallanca, C., Dalessandro, E., Ferraro, F.R., et al.\ 2010, ApJ, 725, 1165 

Papitto, A.; Ferrigno, C.; Bozzo, E. et al., 2013, Nature, 501, 517

Paty, C. \& R. Winglee 2004, Geophys. Res. Lett., 31, L24806

Piotto, G.; Milone, A. P.; Bedin, L. et al., 2015, AJ, 149, 91

Puschnig, J., Hayes, M., {\"O}stlin, G., et al. 2017, MNRAS, 469, 3252

Reed et al. 2016, Proc. AIAA Space 2016. 

Ricker, G. R., Vanderspek, R., Winn, J., Seager, S., et al. 2016, SPIE, 9904, 2

Ricotti, M. \& Gnedin,N. 2005 ApJ, 629, 259

Rodney, S.A., Balestra, I., Bradac, M., et al., 2018, Nature Astronomy, 2, 324

Roberge, A.; Lecavelier des Etangs, A.; Grady, C. et al., 2001, ApJ, 551, 97

Roesler, F. et al. 1999, Science, 283, 353

Roth, L. et al. 2014, Science, 343, 171

Ruiz--Lapuente, P. 2018, arXiv 1812.04977 

Schaefer, B.E. \& Pagnotta, A. 2012, Nature, 481, 164

Siana, B., Teplitz, H.I., Ferguson, H.C., et al.\ 2010, ApJ, 723, 241

Siegmund, O.; 2017, Proc. SPIE, 10397, 11

Sing, D.; Fortney, J.; Nikolov, N. et al., 2016, Natur, 529, 59

Simon 2019 ApJ, 863, 89

Smartt, S., 2009, ARAA, 47, 63 

Smith, B.M., Windhorst, R.A., Jansen, R.A., et al.\ 2018, ApJ, 853, 191

Spake, J.,; Sing, D.; Evans, T. et al. 2018, Natur, 557, 68

Tumlinson, J., Peeples, M.~S., \& Werk, J.~K., 2017, ARA\&A, 55, 389  

Van Dyk, S. A., Li., W., Filippenko, A. V. 2003, PASP., 115, 1. 

Van Dyk, S. A.,  20119, ApJ, 875, 136 

Wang, J.,  2006, Proc. SPIE, 6221, 03 

Weisz et al. 2014 ApJ, 789, 147

Werk, J.K., Prochaska, J.X., Cantalupo, S., et al., 2016, ApJ, 833, 54

Whitaker, K.E., Bezanson, R., van Dokkum, P.G., et al., 2017, ApJ, 838, 19  

Zurek et al. 2016, MNRAS, 460, 3660

Zurek et al. 2009 ApJ, 699, 1113

\end{document}